\documentclass[traditabstract]{aa}
\usepackage{graphicx}
\usepackage{txfonts}
\usepackage{booktabs}
\usepackage[colorlinks=true,linkcolor=blue,citecolor=blue,urlcolor=blue]{hyperref}
\graphicspath{{figures/}}
\newcommand{\fluxunit}{\ensuremath{\mathrm{erg\,cm^{-2}\,s^{-1}}}}

\newcommand{\zenodo}{\url{https://zenodo.org/records/23640281}}
\newcommand{\github}{\url{https://github.com/doroshv/eropngcat}}
\begin{document}
\title{ePNGCAT: the first eROSITA all-sky data release}
\author{Victor Doroshenko}
\institute{}
\titlerunning{ePNGCAT: the first eROSITA all-sky data release}
\abstract{The eROSITA telescope array aboard the joint German-Russian Spektrum-Roentgen-Gamma (SRG) mission delivered the deepest soft all-sky X-ray survey to date scanning the entire sky more than four times between Dec 2019 and Feb 2022. The potential of the survey remains, however, hampered by the fact that only data for the half of the sky owned by the German consortium have been released publicly, whereas the other half remains accessible only to the Russian consortium, effectively making eROSITA a ``half-sky'' survey for most applications. The only publicly accessible all-sky ``data product'' as of 2026 is a single press-release image published in June 2020 jointly by both consortia to celebrate the completion of the first all-sky survey (eRASS1). Here I present the first all-sky eROSITA catalogue based on this image containing around 63 thousand sources with estimated completeness of about 50\% above $3\times10^{-13}$\,\fluxunit\ in the 0.2--2.3\,keV band and positional accuracy of $\sim1\arcmin$ validated against full eRASS1 for half of the sky and several external catalogues. The catalogue is thus comparable with 2RXS, the deepest truly all-sky survey generally available, giving a first glimpse of the eastern Galactic hemisphere with eROSITA. In this hemisphere, about 11\,000 detected sources have no counterpart in ROSAT or other public X-ray catalogues and are expected to be genuine new X-ray sources.
Besides the catalogue, I also release calibrated HEALPix maps to also facilitate studies of large-scale extended emission like eROSITA bubbles or Galactic absorption requiring full sky coverage, available at \zenodo, and the full analysis pipeline for full reproducibility of the results, available at \github.}
\maketitle
\nolinenumbers

\section{Introduction}
\label{sec:intro}
The joint German--Russian Spektrum-Roentgen-Gamma (SRG) mission, carrying the eROSITA X-ray telescope array, was launched on 13 July 2019 \citep{2021A&A...656A.132S,2021A&A...647A...1P} with the primary goal of conducting the first all-sky X-ray survey since ROSAT with major improvements in sensitivity, spatial resolution, and energy coverage. According to the collaboration agreement, data rights were split between the national scientific consortia, each getting half of the sky in Galactic coordinates separated by the meridian passing through the Galactic centre: the Russian consortium exploits half of the sky mostly observable from the Northern hemisphere (eastern), and the German consortium the other half. The survey can now be considered complete, and the German consortium (eROSITA\_DE) has already released most of the data for general access in two data releases. The first included source catalogues, photon lists and source products for the first half-sky survey eRASS1 \citep{2024A&A...682A..34M}, and the second includes data for the first three surveys (eRASS:3) and contains only source catalogues \citep{2026A&A...712A.171R}. 

In the absence of equivalent releases by the Russian consortium, the other half of the sky remains largely inaccessible to the scientific community with the only exception being the press-kit image\footnote{Description and credits: \url{https://www.mpe.mpg.de/7461950/erass1-presskit}.} published jointly in June 2020 to celebrate completion of the first survey. Here I present analysis of this public image aiming to recover an approximate all-sky source catalogue. 
The main motivation for this work is to enable X-ray follow-up of at least the brightest objects in this hidden part of the sky while XMM-Newton and Chandra remain operational. The other reason is my personal wish to bring to a logical conclusion the work I began for the eROSITA bubbles discovery paper \citep{2020Natur.588..227P}, which included detection and removal of compact sources to study the large-scale emission.  

 The following sections describe the input data and processing (Sect.~\ref{sec:methods}), catalogue construction (Sect.~\ref{sec:catalogue}), validation (Sect.~\ref{sec:validation}), and available products (Sect.~\ref{sec:products}). Having left academia two years ago, I intend this work to remain a preprint and catalogue release rather than a proper journal publication, as I have no time or funding to go through the full review process and pay publication fees. The full processing pipeline is, however, released alongside the catalogue (\github; data products at \zenodo) to allow the analysis to be reproduced and improved. 

\section{Input data and image processing}
\label{sec:methods}
\subsection{The image and reference catalogues}
\begin{figure*}[t]
\centering
\includegraphics[width=0.98\textwidth]{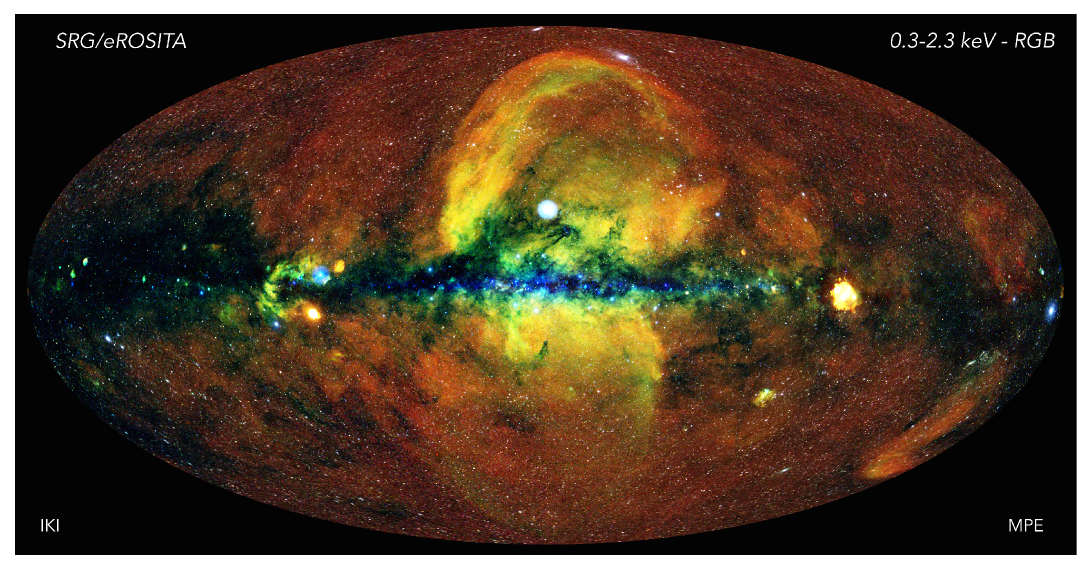}
\caption{Input eRASS1 colour image, reproduced without additional overlays. Red, green, and blue encode the approximate 0.3--0.6, 0.6--1.0, and 1.0--2.3\,keV ranges. The Galactic centre is near the middle, longitude increases towards the left, and the Galactic plane runs horizontally. Compact sources are superposed on diffuse emission, while the brightest objects have clipped cores. The image is a display rendering, so neither brightness nor colour can be interpreted directly as a calibrated flux or spectrum. Original image credits are retained.}
\label{fig:input}
\end{figure*}

The main input for the analysis is a single eight-bit RGB PNG (hence ePNGCAT name) from the official eRASS1 press kit released in June 2020 to celebrate completion of the first all-sky survey and shown in Fig.~\ref{fig:input}. According to the press-release text, the red, green, and blue channels represent 0.3--0.6, 0.6--1.0, and 1.0--2.3\,keV bands, respectively. It is also mentioned that the image is in a Hammer--Aitoff sky projection, and that it was smoothed, although the exact coordinate conventions, intensity transformation, and smoothing kernel are not given. These must therefore be estimated empirically.

The primary calibration and validation reference used below is version 1.2 of the eRASS1 Main catalogue\footnote{\url{https://erosita.mpe.mpg.de/dr1/AllSkySurveyData_dr1/Catalogues_dr1/}.} from eROSITA\_DE DR1 updated in Jan 2026. It uses the same data as the press-kit image, although the processing and energy bands differ. As calibration reference, I use eRASS1 catalogue fluxes in the broad 1B band (0.2--2.3\,keV, labelled 1 in the products), P1 (0.2--0.5\,keV), P2 (0.5--1.0\,keV), and P3 (1.0--2.0\,keV) to calibrate against, and report calibrated image fluxes in these four bands (P1--P3 roughly correspond to the image channels, although none is an exact band match, so the calibration is purely empirical).

Besides eRASS1, I also compare results with ROSAT \citep{1999A&A...349..389V, 2016A&A...588A.103B} and XMM-Newton catalogues, namely the XMM-Newton slew survey XMMSL3 \citep{2008A&A...480..611S} and the serendipitous source catalogue in its 5XMM-DR15 release \citep{2020A&A...641A.136W}\footnote{\url{https://www.cosmos.esa.int/web/esdc/esasky-catalogues} and \url{https://xmmssc.irap.omp.eu/Catalogue/5XMM-DR15/5XMM_DR15.html}}. Note that these generally sample different epochs, so intrinsic variability and band mismatch contribute to their flux differences from eRASS1 and a direct flux comparison is not justified. They nevertheless provide useful cross-checks of the approximate flux scale, astrometry, and source recovery in both halves, not only in the part covered by eRASS1. The ROSAT bright-source catalogue \citep{1999A&A...349..389V} also supplies astrometric references for initial image registration. In addition, the ART-XC \citep{2022A&A...661A..38P, 2024A&A...687A.183S}, MAXI/GSC \citep{2018ApJS..235....7H, 2018ApJS..238...32K}, MAXI/SSC \citep{2016PASJ...68S..32T}, and RXTE slew-survey \citep{2004A&A...418..927R} catalogues provide complementary checks at higher energies (see accompaining pipeline code and full catalogue for more detail). Finally, SNRcat \citep{2012AdSpR..49.1313F} is used to define exclusion masks around known supernova remnants to filter out a large number of spurious sources present both in eRASS1 and detectable in press-kit image that could otherwise contaminate the calibration.

\subsection{Image registration}
As a first step, a coarse registration and projection reconstruction is obtained from the outline of the projected sky ellipse, which fixes its centre, semi-axes, and rotation. The raw all-sky image is then searched for brightest local peaks, and found sources are registered against isolated eRASS1 and ROSAT BSC sources. Here and for most of the analysis described below I divide the full image into two halves corresponding to the German/Russian data-rights split, and in each half fit the registration on a random half of the reference sources, noting offsets for the remaining half to estimate residual misalignment. This also allows one to verify the assumption that the original image is in a Hammer--Aitoff projection. This is most easily done using the part covered by eRASS1 where a Hammer--Aitoff solution gives a median offset of $51\arcsec$ for fitting and about $85\arcsec$ for hold-out sources, compared to $802\arcsec$ for a Mollweide solution hold-outs. Adding low-order polynomial distortion terms to the Hammer--Aitoff solution does not improve it (hold-out medians of 51--$56\arcsec$). 

Applying the same approach to the entire image, however, leaves a median eastern-half offset of $599\arcsec$ against held-out ROSAT bright sources, which suggests that the press-release image is a composite of images independently processed  by each consortium and only roughly aligned in post-processing.
Fitting a separate eastern Hammer--Aitoff solution reduces the offset to $94\arcsec$, i.e. residual offset becomes comparable for both. The eastern solution is necessarily constrained by ROSAT BSC sources alone. Repeating the BSC-only fit in the western half reproduces the eRASS1-based solution to within 0.1 native pixels, which supports this approach. For XMMSL3 sources not used in either fit median offsets of $61\arcsec$ and $69\arcsec$ in the western and eastern halves are detected. 

There also seems to be some misalignment of individual color channels as it was found that in the western half the blue channel requires an approximately one-native-pixel horizontal shift, whereas the fitted eastern shifts are below 0.04 pixels. To conclude, initial registration is about $85\arcsec$ accurate if channels and halves are registered separately. This figure is not to be confused with the overall positional accuracy of the catalogue, as source detection in the all-sky image is obviously not optimal due to projection effects squishing point sources to ellipses especially around the poles. The projection parameters estimated above are only used to define a rough transform of the Hammer--Aitoff-projected original to other projections required for source detection and position refinement downstream on tile level. 

\begin{figure*}[t]
\centering
\includegraphics[width=0.34\textwidth]{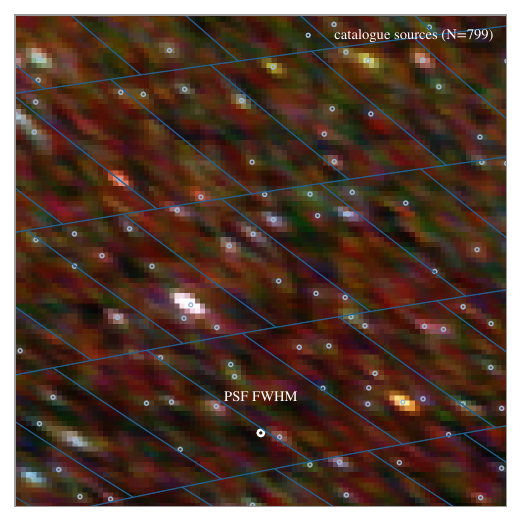}\hfill
\includegraphics[width=0.63\textwidth]{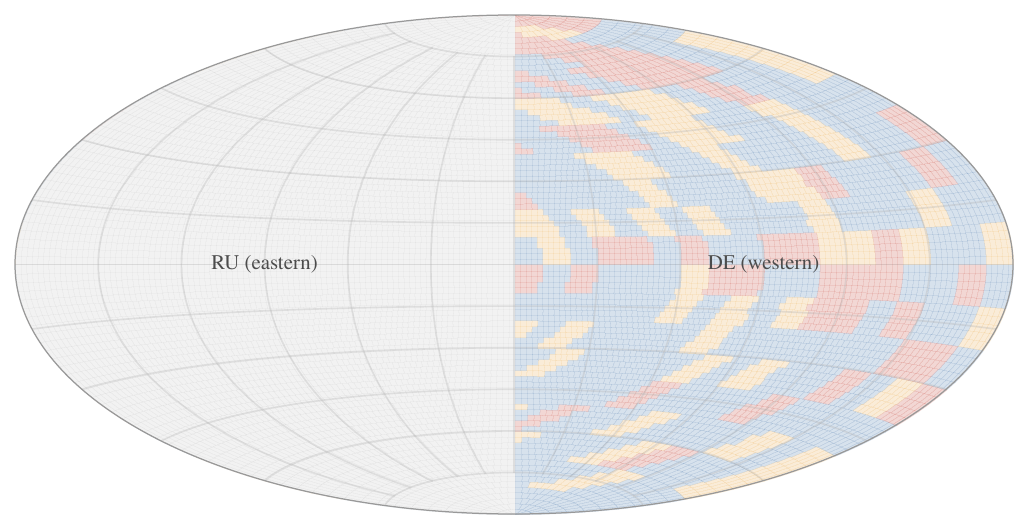}
\caption{Spatial sampling of the image. Left: a roughly $14\degr$ field near $(l,b)=(120\degr,-45\degr)$ with tile boundaries in blue and image-catalogue positions as circles. The labeled white circle indicates the fitted effective sky-term PSF FWHM, approximately $10.7\arcmin$. Right: actual tile footprints in the block-based split of the western (DE) half; blue, orange, and red mark training, validation, and test regions. The eastern (RU, grey) half has no eRASS1 coverage, so no split is possible there and the trained models are only applied. Colours are shown at reduced opacity.}
\label{fig:geometry}
\end{figure*}

Final alignment and further processing in general require proper source detection. To do that, I divide the sky into 10568 tiles with approximately $2\degr\times2\degr$ cores and overlapping margins and re-project the original image to a tangential projection centred on each tile. The tile size was chosen to be small enough for analysis on a tangent plane while retaining neighbouring sources and background for joint fitting. The 5284 western tiles are grouped into approximately $10\degr$ spatial blocks and randomly assigned to training, validation, and test regions with nominal probabilities of 60, 20, and 20\%, giving 3158, 1152, and 974 tiles, respectively (Fig.~\ref{fig:geometry}). Training regions are used to fit a smooth residual position correction (Sect.~\ref{sec:posunc}), the intensity transformation and source models, including the machine-learning relation between image features and catalogue fluxes as described in the next section. Test regions are reserved for final estimates of the resulting detection completeness, association reliability, positional accuracy, and flux accuracy, including the effects of confusion, and are not used in any calibration or model fitting after the global registration. 

For each tile the sources are detected for each tile as local maxima of a filtered image (\texttt{scipy.ndimage.maximum\_filter}) and measured (Sect.~\ref{sec:catalogue}), and unique, robust matches to eRASS1 in the western half are used for tile registration and final position corrections and flux calibration.
This step is i.e. similar to initial coarse alignment. The main difference is that instead of point sources distorted by full-sky projection, more round and better constrained tile-positions are used, and at this stage we already have an estimate of coarse position alignment uncertainty, so candidate matches may be selected more reliably. With matches identified, additional global polynomial coordinate distortion model is fit to all sources. 
When doing so, the test-validation scheme described above and mainly used for final accuracy assessement, is not respected, i.e.  same source-level random hold-out as for coarse alignment rather than these spatial regions is used. As a consequence, a fraction of sources used for alignment lie in test regions, i.e. train/test split is not perfect. With only a few global projection parameters fitted to thousands of sources, this has, however, a negligible effect on the test statistics and final estimates of positional uncertainty. Once global projection parameters estimated, those are applied to all detected sources (and not only those used for alignment) to convert image positions to sky coordinates. The residual uncertainty is then estimated using hold-out tiles as described in sections \ref{sec:posunc} and \ref{sec:astrometry}.

\subsection{Intensity mapping and PSF}

Tile-based source detection is also used to estimate initial intensity scaling for the image, i.e. revert intensity scale transformation applied to raw eROSITA images to get estethically pleasing press-kit image. To estimate the intensity transformation, I select isolated, unsaturated eRASS1 sources in the training regions and compare their reference fluxes with background-subtracted image peak intensities. Parameters of linear, power-law, exponential, and hyperbolic-sine transformations are fitted on a random half of this sample, and the family with the smallest robust scatter in the flux-to-intensity ratio on the remaining half is then the likely transformation applied to original image. Among tested options, the hyperbolic-sine transformation, the inverse of an asinh display stretch, appears to be such a transform:
\begin{equation}
 I_c=\sinh\left(\frac{v_c-v_0}{A_c}\right),\label{eq:stretch}
\end{equation}
where $v_c$ is the display value in channel $c$, $A_c\simeq100,138,163$ for red, green, and blue, and the adopted common offset is $v_0=-6$. 

\begin{figure*}[t]
\centering
\includegraphics[width=0.98\textwidth]{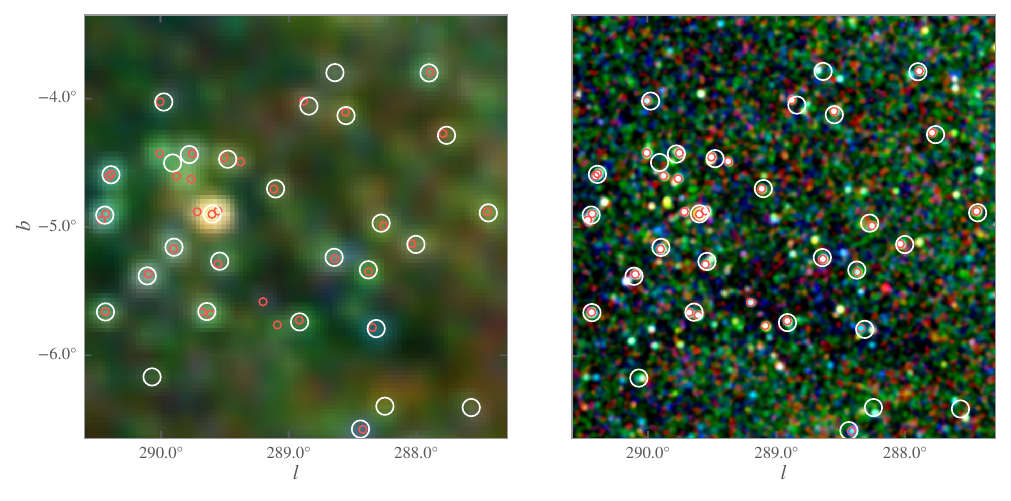}
\caption{A $3.3\degr\times3.3\degr$ test-region tile centred on $(l,b)=(288.9\degr,-5.0\degr)$. Left: press image resampled to the tile grid. Right: eRASS1 DR1 HiPS rate maps in the 0.2--0.5, 0.5--1.0, and 1.0--2.0\,keV bands (red, green, blue) on the same grid, smoothed with a $1\arcmin$ Gaussian, converted back to image intensities with the inverse of the map calibration (Sect.~\ref{sec:products}), and displayed with the press-image stretch of Eq.~(\ref{eq:stretch}), so that both panels share one colour scale. Small red circles mark eRASS1 sources above the 50\% completeness limit $F_{50}=3.12\times10^{-13}$\,\fluxunit; large white circles mark ePNGCAT sources. The panels illustrate the main detection limitations. Close eRASS1 pairs merge into a single detection (4 of 29 ePNGCAT sources). Sources above $F_{50}$ are missed at the tuned detection threshold (7 of 35). Sources below $F_{50}$ are still detected (3), because $F_{50}$ is a sky-averaged value and the limit in an individual tile differs from it. Two detections have no eRASS1 counterpart; one is flagged as extended.}
\label{fig:hires}
\end{figure*}

As with the projection estimate, the resulting intensities are intermediate image units and are not used directly as final fluxes. This empirical transformation only reduces the initial display non-linearity but neither reconstructs photon counts nor restores clipped information.
In particular, flux calibration is also affected by smoothing applied to the original image. I estimate the effective point-spread function (PSF) by fitting elliptical Gaussian profiles to isolated, unsaturated sources. Their covariance matrices are modelled as the sum of an isotropic sky term and a native-pixel smoothing term transformed by the local projection Jacobian. The fitted sky term has $\sigma=4.55\arcmin$, equivalent to a Gaussian full width at half maximum (FWHM) of $10.7\arcmin$ before the additional pixel term. This is the measurement kernel. A separate calibration for scene subtraction adjusts the sky width to null stacked central residuals on 120 training tiles, giving $\sigma\simeq5.74\arcmin$. Both widths characterise the press-kit image, not the instrumental eROSITA PSF, but are the values we are actually interested in and use for source detection and PSF photometry of all sources. The resulting loss of information is illustrated in Fig.~\ref{fig:hires}, which compares one test-region tile in the press image with the eRASS1 rate maps of the same field: of 694 eRASS1 sources in the field, 35 are brighter than the 50\% completeness limit derived in Sect.~\ref{sec:validation}, and the image catalogue contains 29 sources, 27 of which have eRASS1 counterparts.

\section{Source detection and catalogue construction}
\label{sec:catalogue}

\subsection{Detection and bright-source treatment}
At this point the rough geometric and intensity transformations of the original image to linearized form are determined and can be used for source detection and photometry on extracted tiles. Final source detection begins with an estimate of the local background and pixel scatter in each linearised channel. After background subtraction, the channels are combined with noise-based weights and filtered with a kernel matched to the effective source profile obtained above to create an optimal ``detection'' image. Weights are learned as part of the process by optimizing detection performance on validation regions. 

Local maxima above a threshold (also calibrated on validation sample) in this filtered image become source candidates. I compare isotropic and projection-dependent filters and tune their parameters on training regions in order to optimize detection performance. Here training regions are used to set parameters and validation regions select the configuration using $U=N_{\rm real}-4N_{\rm spur}$, where reference associations, corrected for chance matches, estimate the numbers of real and spurious detections. Maximising $U$ is equivalent to keeping detections whose marginal reliability is at least 0.8. The penalty limits contamination by diffuse structure without selecting only the brightest objects. The validation optimum is at signal-to-noise ratio (S/N) 4; I adopt a lower threshold of 3.5 to retain more fainter candidates, accepting the resulting loss of reliability. For comparison, a baseline detection on the raw image luminance with a $5\sigma$ threshold reaches 50\% completeness at $5.8\times10^{-13}$\,\fluxunit\ on the validation regions, whereas the tuned detection reaches $3.1\times10^{-13}$\,\fluxunit, i.e. it is about factor of two deeper than naive. This S/N is an empirical contrast statistic for the rendered image, not an eROSITA-like detection likelihood or a Gaussian false-alarm probability (the empirical reliability as a function of S/N is given in Sect.~\ref{sec:validation} instead).

For each candidate, the pipeline fits a common position and separate channel amplitudes and backgrounds, including nearby detected sources in the fit. Measurements are made both on the sampled tangent-plane image and on the original pixels, with fixed and variable profile widths. Background-subtracted aperture sums within $6\arcmin$ and $10\arcmin$ provide additional measures of source brightness (i.e. features for flux regression model). These amplitudes and aperture sums are still in linearised image units; their conversion to physical flux is learned from eRASS1 as described below.

Bright sources are saturated and lose information about the peak intensity, so special treatment is needed to recover the flux using wings of the PSF. Connected regions of clipped pixels in the centre are therefore introduced as candidates even when ordinary peak detection fails. For each such region, the pipeline fits the surrounding emission with a radial sigmoid profile (rather than a Gaussian), with a shared centroid and separate amplitudes and radii for each channel. The motivation here is that a source might be saturated only in one of the channels, or saturated to a different degree, so the degree of ``clipping'' might differ. An initial fit removes the broad component before searching for fainter neighbour sources in PSF wings. These are masked out and the bright-source fit is refined to get a robust starting point for the final joint fit including the bright source, blended fainter sources, and a smooth background. The parameters of the sigmoid are not used to directly measure image flux for bright source, but rather as features for the final flux model.

Considering the relatively low resolution of the input image, extended sources can mimic a saturated point source in this approach. Introduction of the sigmoid component described above is, therefore, not done for known SNRs and galaxy clusters, and also in regions with very bright diffuse backgrounds (e.g. knots in the eROSITA bubbles and siilar regions). The final sigmoid parameters are also rejected if the fitted amplitude, radius, or effect on the residual image is implausible and in this case detection falls back to single core model. These fixed safeguards were chosen from visual inspection of bright point sources, SNRs and galaxy clusters and diffuse fields, including Cyg~X-1, and inspection of the amplitude-area and radius-area relations measured across both image halves. They are empirical limits defining where modes of the trained flux model are applicable, and defined on manual inspection of problemmatic cases with no independent classifier validation. The examples in Appendix~\ref{app:bright} illustrate how both models describe point sources but do not establish improved neighbour recovery or preservation of diffuse flux. These corrections apply to the 2\,759 sources with clipped pixels (``saturated'' in Table~\ref{tab:summary}, 4\% of the catalogue). Clipping in the rendered image is not restricted to the brightest X-ray sources: the median broad-band flux of saturated sources is $4.4\times10^{-12}$\,\fluxunit, and 10\% are fainter than $8\times10^{-13}$\,\fluxunit. These are mostly sources overlapping bright extended emission, which is artificially enhanced by the smoothing and the non-linear intensity stretch of the input image. Of these, 874 are also flagged as possibly extended. Their fluxes and positions should be treated with caution, although they're probably of little interest as most of the affected sources are bright and thus well known so information from other catalogs is available. 

\subsection{Position uncertainty calibration}
\label{sec:posunc}

Using the features obtained as described above (i.e. centroid positions for the optimised PSF model), I estimate residual misalignment and positional uncertainty for calibrated positions. The latter might be, and apparently is, a function of detection significance. To estimate this dependence, I compare positions from the filtered detection image, local intensity centroids, profile fits, and joint scene fits with a smooth spline background. Residual corrections in Galactic coordinates are fitted on training sources (in the western half only; eastern positions retain the BSC-based registration without such a correction, which contributes to their larger offsets in Sect.~\ref{sec:external}), and validation regions select either a single estimator or an S/N-dependent combination. 

The base positional error model is calibrated on validation associations. The Rayleigh scale of the offsets is measured in ten quantile bins of S/N, interpolated in log S/N, and required not to increase with S/N. It falls from $84\arcsec$ at S/N$\,\approx3.7$ to $25\arcsec$ at S/N$\,\approx22$ and is held constant outside this range. A parametric form $\sigma^2=a^2+(b/{\rm S/N})^2$ cannot follow the steep drop where the selected estimator switches from centroids to profile fits, and it overestimated the uncertainties in the brightest bins by a factor of about 1.5. Saturated detections receive a separate constant error of $19.7\arcsec$. These are per-axis uncertainties; for an isotropic two-dimensional Gaussian, the 68\% containment radius is approximately $1.51\sigma_{\rm pos}$. The selected combination mainly uses centroids for faint detections and native-pixel profile fits for brighter ones. The result is used as an estimate of positional uncertainty for the entire sky as the eastern half lacks eRASS1 reference, with an additional eastern term described in Sect.~\ref{sec:external}.

\subsection{Flux calibration}

To obtain final flux estimate, a separate model is trained and validated on sources detected in individual tiles, separately for the training, validation, and test regions. Inputs include the channel profile amplitudes and aperture sums from the fit described above, local backgrounds and pixel scatter, profile widths and fit residuals, saturation area, bright-source profile parameters, S/N, neighbour counts and separations, local projection area and distortion, and absolute Galactic latitude. The model is i.e. data driven and predicts the logarithm of the reference energy flux from the measured image properties, with one model for each output band, which is then compared with ground truth on validation sample to assess accuracy of the empirical flux calibration model. 
Several models were tested: weighted-amplitude calibrations, ridge regression, and gradient-boosted decision trees (\texttt{scikit-learn} \texttt{HistGradientBoostingRegressor} and \texttt{XGBoost}, with hyper-parameters tuned on an internal split of the training sample). The \texttt{HistGradientBoostingRegressor} provided the best performance on the validation sample and thus was adopted. Each band is fitted separately on training sources using that family.

Reference associations initially use the brightest eRASS1 counterpart within $4\arcmin$ above a broad-band flux of $10^{-13}$\,\fluxunit. Only eRASS1 point sources were considered for calibration (i.e. \texttt{EXT\_LIKE}=0), and only image detections having a final separation below $150\arcsec$ from eRASS1 counterpart, and a positive flux in the fitted band were considered. For these robust associations a gradient-boosted regressor was trained on training regions and validated on validation regions. Flux uncertainty was estimated by training separate quantile regressors for the 16th and 84th percentile bounds and compared with validation scatter of the main model prediction to ensure both are consistent. 

Once positions and fluxes are calibrated, additional systematic error terms might be needed both for positions and fluxes. To estimate these, unique one-to-one associations within $4\arcmin$ of point-like eRASS1 sources, requiring both image and reference broad-band fluxes above the validation 50\% completeness limit, $3.08\times10^{-13}$\,\fluxunit, were considered (this is measured on the validation regions; the test-region value of $3.12\times10^{-13}$\,\fluxunit\ quoted in Sect.~\ref{sec:validation} is reserved for the final performance assessment). Additive and fractional error terms are fitted on training regions and compared on validation regions with a Bayesian-information-criterion penalty for additional parameters. The training-fitted positional correction defines the two-sigma association cut used for this flux-error calibration. Selected terms are applied unchanged to the catalogue and evaluated on test regions. An additive $22.6\arcsec$ term is selected for positions; fractional terms of 7.6, 4.7, and 6.2\% are selected for the broad-band, P1, and P3 flux intervals, respectively. Each is added in quadrature to the corresponding base error, with the fractional term proportional to the source flux. P2 intervals remain unchanged. Table~\ref{tab:flux} compares base and corrected coverage on identical test associations.

\section{Catalogue properties and validation}
\label{sec:validation}
\subsection{Detection statistics and performance}
\begin{figure}[t]
\centering
\includegraphics[width=0.48\textwidth]{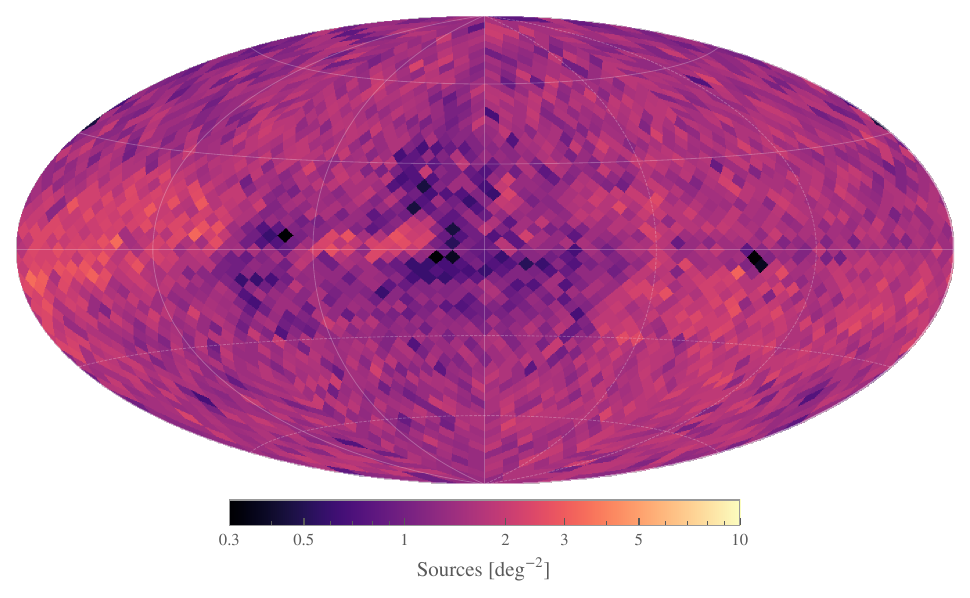}
\caption{Sky distribution of all detected sources in Galactic coordinates, with the Galactic centre in the middle and longitude increasing to the left. Shown is the surface density in equal-area HEALPix cells with $N_{\rm side}=16$, each covering approximately $13.4\,\mathrm{deg}^{2}$; the logarithmic colour scale gives sources per square degree. Flagged sources are included. Spatial variations combine real source structure, diffuse-emission contamination, and the non-uniform selection of the rendered image.}
\label{fig:sky}
\end{figure}

The final catalogue contains 63\,260 sources, with 32\,379 in the western and 30\,881 in the eastern half (Table~\ref{tab:summary}). Throughout, the \emph{western footprint} ($180\degr\lesssim l<360\degr$, labelled DE in the figures and products) is the half of the sky covered by public eRASS1 data; it is also referred to as the reference footprint, since every source there can be checked against eRASS1. The \emph{eastern footprint} ($0\degr\lesssim l<180\degr$, labelled RU) is the other Galactic half, for which no public eROSITA data exist and the image catalogue is the only eROSITA-based source list. The scientific data-rights boundary is at Galactic longitudes approximately $179.944\degr$ and $359.944\degr$, through the Galactic poles and Sgr~A*, rather than exactly $180\degr$ and $0\degr$. This convention is distinct from the fitted join between the two image halves. The sky distribution and surface density are shown in Fig.~\ref{fig:sky}. Quality flags identify potentially problematic entries so that users can select samples suited to their applications.

\begin{table}[t]
\caption{Content of the ePNGCAT catalogue.}
\label{tab:summary}
\centering
\begin{tabular}{lr}
\toprule
Quantity & Value \\
\midrule
All sources & 63\,260 \\
Western footprint & 32\,379 \\
Eastern footprint & 30\,881 \\
Saturated sources & 2\,759 \\
Possibly extended sources & 9\,849 \\
Sources within SNRcat masks & 1\,665 \\
Sources flagged near image edges/joins & 455 \\
Excluded sky area & $188\,\mathrm{deg}^{2}$ \\
Excluded sky fraction & 0.46\% \\
\midrule
Unmatched candidates, eastern footprint$^{a}$ & 13\,983 \\
\quad expected real$^{a}$ & ${\approx}10\,800$ \\
\bottomrule
\end{tabular}
\tablefoot{Flag categories overlap. ``Possibly extended'' refers to the image profile or saturated area and is not a physical source classification. $^{a}$~Outside SNRcat masks and $0.5\degr$ from the image joins, without counterparts within three combined positional errors in ROSAT, XMMSL3, 5XMM, or ART-XC catalogues (Sect.~\ref{sec:orphans}).}
\end{table}

In the spatial test regions, excluding remnant masks, 5798 detections yield a chance-corrected association reliability of 87.5\% (i.e. when comparing with eRASS1). Chance associations are estimated by displacing positions by $15\arcmin$ and repeating the match. Since eRASS1 is much deeper than the image, a detection without an eRASS1 counterpart is almost certainly not a genuine point source, so this reliability is effectively the purity of the catalogue: roughly one detection in eight is spurious. In practice these are mostly fragments of diffuse or extended emission and residual structure around bright sources, rather than noise peaks which can be expected to be smoothed out by smoothing applied to press-kit image. For point-like reference sources, completeness reaches 50\% at $3.12\times10^{-13}$\,\fluxunit\ in 0.2--2.3\,keV. The monotone completeness estimate plateaus near 87\%, so a more commonly used 90\% limit cannot be assigned. The 50\% limits range from $2.51\times10^{-13}$ to $3.79\times10^{-13}$\,\fluxunit\ across the tested latitude intervals. These values are obviously worse than the underlying survey sensitivity because the input image is degraded; they characterise the degree of degradation and the robustness of the recovery procedure rather than the actual eROSITA sensitivity. Nevertheless, similar figures shall apply also to eastern half of the sky, although here independent validation is not possible. 

\begin{figure*}[t]
\centering
\includegraphics[width=0.47\textwidth]{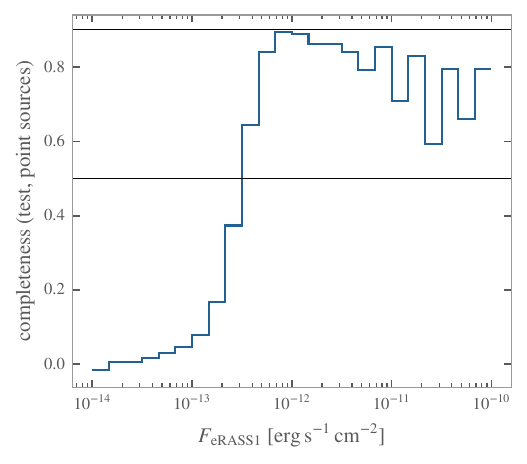}\hfill
\includegraphics[width=0.47\textwidth]{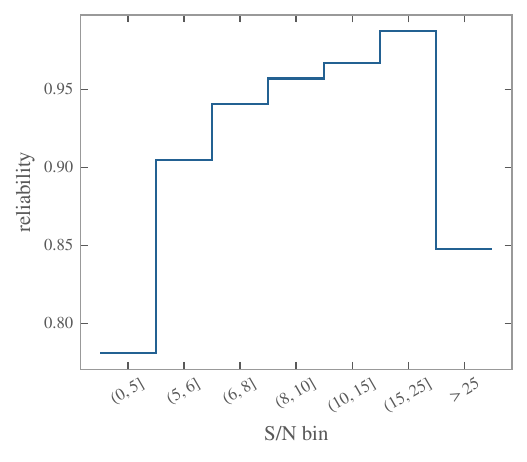}
\caption{Base-model detection validation on the spatial test regions, excluding known remnant regions. Left: point-source completeness against eRASS1 as a function of reference 0.2--2.3\,keV flux, averaged over the test footprint. Right: chance-corrected association reliability in S/N bins. The final bin also contains saturated detections, whose formal detection significance is represented as infinite in the catalogue. These detections should not be interpreted using ordinary noise statistics. The 90\% completeness level is not reached by the monotone completeness estimate.}
\label{fig:detection}
\end{figure*}

Reliability is 78.1\% below S/N 5 and 98.8\% at S/N 15--25, compared with 84.8\% in the highest bin, which also includes saturated sources (Fig.~\ref{fig:detection}). Brightness therefore does not guarantee an unambiguous point-source association. Saturation, blends, and structured emission can all invalidate that interpretation and so sources flagged as saturated must be treated with caution. These are, however, either parts of extended emission detected as point sources or well-known objects, so their reported fluxes and positions are of little interest anyway. 

\subsection{Astrometric and photometric accuracy}
\label{sec:astrometry}
The accuracy estimates in this subsection use only eRASS1 point sources in the western spatial test regions as this largely avoids the inter-epoch variability that would be inevitable if ROSAT and XMM-Newton catalogues were used. Unfortunately, eRASS1 only covers half of the sky, so estimates below might be considered optimistic for the eastern half (see Sect.~\ref{sec:external} for a direct comparison).  

The 4380 matched point sources in the test subset have median, 68th-percentile, and 90th-percentile separations of 72, 105, and $176\arcsec$. The offsets depend strongly on S/N: the median decreases from $99\arcsec$ below S/N 5 to $34\arcsec$ at S/N 15--25 and $23\arcsec$ above S/N 25. This is slightly better than the initial alignment obtained on the full image, but of course still limited by input image resolution and smoothing. 
About 40 and 79\% of offsets fall within one and two combined per-axis base uncertainties. The corresponding fractions after the global correction are 45 and 82\%, compared with 39 and 86\% for a circular two-dimensional Gaussian. The containment fractions refer to radial offsets measured in per-axis error units; they are not one-dimensional Gaussian coverage probabilities.

For flux validation I additionally require a separation below $150\arcsec$ and a positive reference flux in the relevant band, as mentioned above. Table~\ref{tab:flux} summarises the results. The broad-band multiplicative scatter is 27\%, corresponding to a robust width of approximately $-21\%$ to $+27\%$ around the median flux ratio. The softer P1 band is less well constrained, as expected from the band mismatch and sensitivity to spectral shape. The nominal 16--84\% base intervals contain 67--69\% of the reference fluxes; the corrected intervals contain 68--71\% on the same samples. Note that the figures reported come from direct comparison of fluxes reported in eRASS1 and recovered from the public image for \textit{unseen} parts of the sky, and thus can be considered robust. 

\begin{table*}[t]
\caption{Flux validation and interval coverage on matched point sources in the spatial test regions.}
\label{tab:flux}
\centering
\begin{tabular}{lcrrrrrr}
\toprule
Band & Energy (keV) & $N$ & Bias (\%) & Scatter (\%) & Outliers (\%) & $C_{\rm base}$ (\%) & $C_{\rm corr}$ (\%) \\
\midrule
1 & 0.2--2.3 & 3697 & $-0.4$ & 26.5 & 4.0 & 66.8 & 71.3 \\
P1 & 0.2--0.5 & 3471 & $+1.8$ & 63.1 & 20.7 & 67.5 & 68.1 \\
P2 & 0.5--1.0 & 3693 & $+0.1$ & 31.7 & 8.4 & 69.4 & 69.4 \\
P3 & 1.0--2.0 & 3680 & $-1.3$ & 38.2 & 10.2 & 67.8 & 69.9 \\
\bottomrule
\end{tabular}
\tablefoot{For the flux ratio $R=F_{\rm image}/F_{\rm eRASS1}$, let $m=\mathrm{median}(\ln R)$ and $s=1.4826\,\mathrm{median}|\ln R-m|$. Bias and multiplicative scatter are $100(\exp m-1)$ and $100(\exp s-1)$, respectively. The latter describes the upward width of a robust multiplicative interval; the downward width is $100(1-\exp[-s])$. It is not a symmetric percentage error or a measured containment interval. Outliers have $R/\exp m<0.5012$ or $>1.9953$, approximately a factor of two from the median ratio. The last two columns give coverage of the base and corrected 16--84\% prediction intervals on the same associations. All rows require separation below $150\arcsec$ and positive reference flux in the relevant band.}
\end{table*}

The small overall bias conceals a flux dependence. In the faintest populated broad-band interval, $10^{-13}<F_{\rm eRASS1}\leq3.16\times10^{-13}$\,\fluxunit, the median image flux excess is 18.3\%. This is consistent with preferential selection of upward fluctuations near the detection limit. Conversely, successful flux measurements of saturated matches do not establish complete recovery of saturated sources: the validation sample already requires a sufficiently close association and no blending. These selection effects matter when comparing the quoted scatter with other catalogue measurements. 

\subsection{External comparisons and number counts}
\label{sec:external}
For external point sources brighter than $2\times10^{-12}$\,\fluxunit\ in their native comparison bands, the image catalogue recovers 79 and 78\% of XMMSL3 sources in the western and eastern footprints, and 90 and 86\% of ROSAT bright sources. Median offsets are approximately $28\arcsec$ in the west and $46\arcsec$ in the east for both catalogues. Note that one can not expect 100\% recovery due to variability of vast majority of sources. The western value is consistent with the eRASS1 offsets of the similarly bright, high-S/N sources in Sect.~\ref{sec:astrometry}; the much larger median in the eRASS1 test sample is driven by its numerous faint detections. The eastern offsets are, however, about 1.7 times larger at the same flux cut in both catalogues, whereas for the fainter 2RXS sources the medians are $49\arcsec$ and $59\arcsec$, a ratio of only 1.2. This is the signature of an additional error that does not scale with S/N rather than of a proportional one. The quadrature difference of the eastern and western Rayleigh scales of the bright-source offsets, in which the errors of the external catalogues cancel, gives about $32\arcsec$ per axis for both XMMSL3 and the BSC. This term is added in quadrature to the positional errors of all eastern sources (column \texttt{POS\_ERR\_ADD}). It reflects the missing eastern residual correction (Sect.~\ref{sec:posunc}) and the broader eastern source profiles discussed below. Only a small part of it is a correctable registration error: the offsets of neighbouring eastern sources to 2RXS are correlated at a level of only about $8\arcsec$ per axis, and neither smooth nor local corrections fitted against 2RXS reduce the offsets of held-out sources. With the additional term, the offsets normalised by the catalogue errors are distributed alike in both halves (Sect.~\ref{sec:orphans}). The multiplicative flux scatter is 66--88\% under the definition in Table~\ref{tab:flux}. It includes variability and band differences and cannot be attributed entirely to image measurement errors. ROSAT sources also participate in global registration, so their comparison is not a wholly independent astrometric test (the eRASS1 figures should be considered more robust, as they are estimated from truly unseen data). Considering that exact processing details for press-kit image are not known, the described uncertainties are just assumed to be empirical and folded into the final catalog. 

The combined bright-source comparison gives a relative eastern-minus-western flux-scale difference of $-11.3_{-2.6}^{+2.7}$\% before correction. The quoted uncertainty is the transformed bootstrap standard error. Catalogue assembly therefore multiplies eastern fluxes and their bounds by 1.127. A subsequent flux-limited 2RXS comparison gives a residual relative difference of $+1.9_{-3.3}^{+3.5}$\% for $F_{\rm 2RXS}>3\times10^{-12}$\,\fluxunit. This restriction reduces differential selection effects but does not establish identical completeness. In a band-matched 0.2--2.0\,keV comparison with 5XMM, 814 associations within $45\arcsec$ have a median flux excess of 2.4\% and multiplicative scatter of 76\%. This supports the approximate scale for closely matched bright sources; it is not an unbiased census of the sky covered by pointed observations.

At higher energies, the most relevant comparison is the first-year ART-XC all-sky survey ARTSS1 \citep{2022A&A...661A..38P}, which covers the same period as eRASS1. The image catalogue recovers 67 and 72\% of its 4--12\,keV sources in the western and eastern footprints, compared with 84\% for eRASS1 in the western footprint; the four-year ARTSS4 \citep{2024A&A...687A.183S}, which is less contemporaneous, gives 70 and 74\% (91\% for eRASS1). ARTSS1 lists only 815 sources in total, whereas the image catalogue contains 63\,260, and purity or ARTSS1 sample is likely lower on average and is not independently verified. The bands differ, so this mostly reflects the much higher soft-band sensitivity of eROSITA, which even the degraded image retains. The image catalogue further recovers 73 and 71\% of the MAXI/GSC sources, and 81 and 76\% of the RXTE slew-survey sources, compared with about 90\% for eRASS1 in the western footprint. These sources are bright above 4\,keV but not necessarily in the soft band; the similar recovery fractions in the two halves nevertheless indicate no significant difference in completeness for bright sources between them. The MAXI/SSC catalogue is shallower but its 0.7--7\,keV band overlaps ours more closely; the image catalogue recovers 74 and 80\% of its 74 western and 61 eastern sources, compared with 92\% for eRASS1, again with no significant difference between the halves given the small samples. Unidentified ART-XC sources are rarely recovered (20 and 8\% for ARTSS1, 30 and 25\% for ARTSS4, compared with 24 and 46\% for eRASS1), consistent with many of them being heavily absorbed or Galactic-plane sources faint below 2\,keV.

\begin{figure*}[t]
\centering
\includegraphics[width=\textwidth]{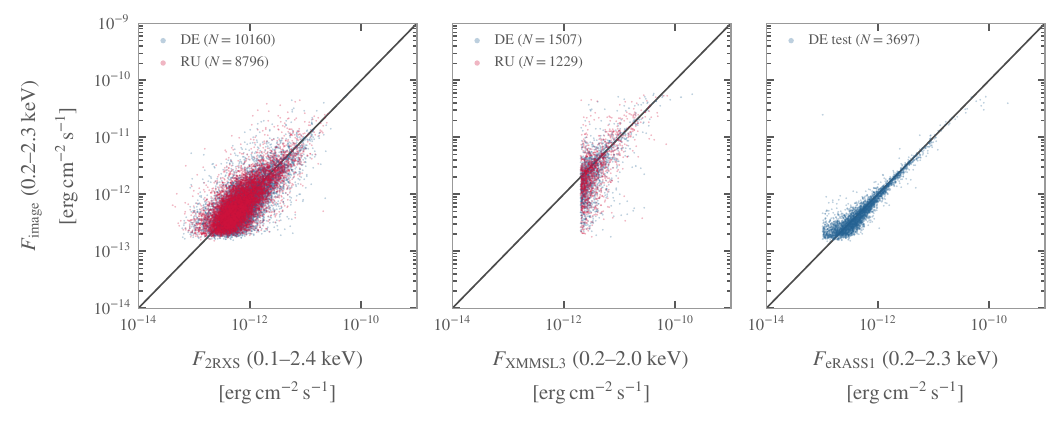}
\caption{Direct flux validation against 2RXS (left), the XMM-Newton slew survey XMMSL3 (middle), and eRASS1 (right). Dots mark individual associations: blue denotes the western footprint (DE) and red the eastern footprint (RU). Image fluxes are in 0.2--2.3\,keV; reference bands are labelled on each abscissa, and the solid lines indicate numerical equality rather than a spectral conversion. The external panels use the final catalogue with the eastern flux-scale correction applied. The 2RXS sample requires detection likelihood at least 10, screening flag zero, and no significant extent; nearest image matches lie within $120\arcsec$. Its native 0.1--2.4\,keV fluxes are estimated from count rates and hardness ratios. The XMMSL3 panel uses point-like detections grouped by unique source with median flux above $2\times10^{-12}$\,\fluxunit\ in 0.2--2.0\,keV and matches within $180\arcsec$. Both external samples exclude remnant masks and a $0.5\degr$ region around the image joins. The eRASS1 panel contains 3697 point-like reference associations in the western test regions, with positive broad-band reference flux and separation below $150\arcsec$, as in Table~\ref{tab:flux}. It uses base-model fluxes. No source-by-source band conversion or variability correction is applied, and the plotted associations are not statistically corrected for chance matches.}
\label{fig:flux-comparison}
\end{figure*}

Figure~\ref{fig:flux-comparison} shows direct flux comparisons with 2RXS, XMMSL3, and eRASS1. The external panels use the final image catalogue, including the eastern flux-scale correction, whereas the eRASS1 panel uses the same held-out base-model sample as Table~\ref{tab:flux}. The tighter relation with eRASS1 is consistent with calibration against that survey and a common observing epoch. The sharp left boundary in the XMMSL3 panel reflects the imposed reference-flux cut, and the flattening of the faint image fluxes illustrates selection near the image detection limit. These are matched samples rather than complete flux-selected populations, so the comparisons are consistency checks only; nevertheless, both external comparisons show a good correlation and rough agreement of absolute flux scale on average.  

\begin{figure*}[t]
\centering
\includegraphics[width=0.47\textwidth]{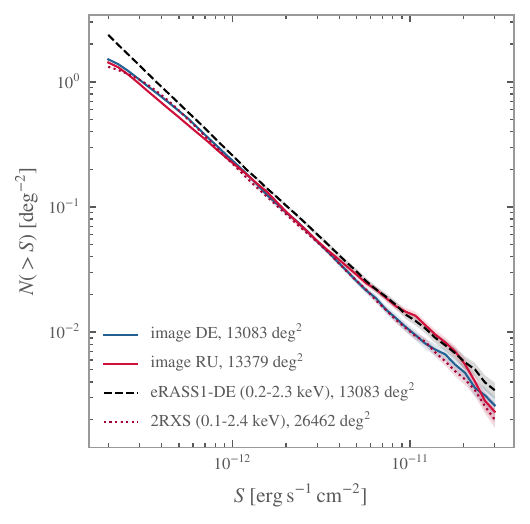}\hfill
\includegraphics[width=0.47\textwidth]{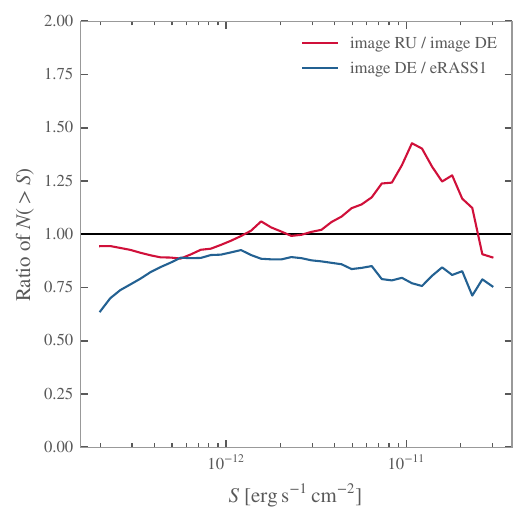}
\caption{Cumulative number counts at $|b|>20\degr$. Left: image catalogues in the western (blue) and eastern (red) footprints, eRASS1 (black dashed), and 2RXS (dotted). Shading shows counting errors $\sqrt{N}/\Omega$, without completeness or calibration uncertainties. Image and eRASS1 fluxes refer to 0.2--2.3\,keV; 2RXS retains its native 0.1--2.4\,keV band. Right: eastern/western image counts (red) and western image/eRASS1 counts (blue). All samples use the same HEALPix mask for the rendered footprint, image-edge margin, $0.5\degr$ join exclusions, and all SNRcat masks; the retained mask area normalises each curve. The 2RXS sample additionally retains its catalogue likelihood and screening cuts. No completeness or band-conversion correction is applied.}
\label{fig:counts}
\end{figure*}

Figure~\ref{fig:counts} compares cumulative source densities above $|b|=20\degr$. All samples use a common geometric mask for the rendered footprint, image edges, joins, and remnant regions. The mask is discretised at HEALPix $N_{\rm side}=512$ and used both to select sources and to compute the retained areas: 13083 and 13379\,deg$^2$ in the western and eastern halves, respectively. These are observed counts without completeness corrections. ROSAT retains its native band, so the curves are consistency diagnostics rather than a common-band population measurement. Between $10^{-12}$ and $3\times10^{-12}$\,\fluxunit\ the western image counts are within about 12\% of eRASS1, and the ratio falls to 0.87 and 0.64 at $5\times10^{-13}$ and $2\times10^{-13}$\,\fluxunit\ as incompleteness sets in. Over the whole range from $3\times10^{-13}$ to $10^{-11}$\,\fluxunit\ the image counts agree with those of 2RXS within about 6\%, i.e. at high Galactic latitude the image catalogue reaches a source density comparable to that of 2RXS. The eastern counts agree with the western ones within 4\% between $10^{-12}$ and $2\times10^{-12}$\,\fluxunit, but are about 10\% lower between $3\times10^{-13}$ and $5\times10^{-13}$\,\fluxunit\ despite the eastern flux-scale correction, which suggests a somewhat lower completeness near the detection limit in the eastern half. Above $5\times10^{-12}$\,\fluxunit\ the eastern counts are higher, but this is based on fewer than 200 sources per half and includes extended structures detected as point sources.

The 2RXS comparison independently confirms the lower eastern completeness near the limit: the image catalogue recovers 39 and 76\% of the western 2RXS sources with 0.1--2.4\,keV fluxes of $3\times10^{-13}$--$10^{-12}$ and $10^{-12}$--$3\times10^{-12}$\,\fluxunit, respectively, compared with 34 and 71\% in the east, while the fractions agree above $3\times10^{-12}$\,\fluxunit. Image-only diagnostics, computed identically for both halves without any reference catalogue, point to the likely cause. Unsaturated eastern sources are on average 13\% broader than western ones (median free-width profile scale 1.07 versus 0.95), and the eastern source density at $|b|>10\degr$ is about 8\% lower. Broader profiles reduce the peak contrast of faint sources and degrade their centroids, consistent with both the lower completeness and the larger positional offsets in the east, and suggest that the two halves of the press image were rendered with slightly different smoothing. The image-only number counts also imply an eastern-minus-western flux-scale difference of $-9.6$\% before correction, consistent with the $-11.3$\% derived from the bright-source comparisons above.

\subsection{Candidates without external counterparts}
\label{sec:orphans}
After exclusion of known SNRs and a narrow region between the two halves of the sky, 61\,428 sources remain, and are matched against the 2RXS, ROSAT BSC, XMMSL3, 5XMM, and ART-XC (ARTSS1 and ARTSS4) catalogues. A counterpart is accepted if it lies within three combined per-axis positional errors, $r<3\,(\sigma_{\rm pos}^2+\sigma_{\rm ext}^2)^{1/2}$, where $\sigma_{\rm pos}$ is the catalogue error, including the eastern term (Sect.~\ref{sec:external}), and $\sigma_{\rm ext}$ is the per-axis error of the external catalogue, measured from its offsets to eRASS1 in the western footprint. These are 8--$17\arcsec$ for 2RXS depending on detection likelihood, $9\arcsec$ for the BSC, $6\arcsec$ for XMMSL3, $4\arcsec$ for 5XMM, and 6--$8\arcsec$ for ART-XC; the quoted errors are unavailable or underestimated by factors of 1.2--3. Unlike a fixed radius used for initial flux comparisons, this criterion treats both footprints alike. It retains 90--97\% of the true eRASS1 counterparts in every S/N bin, and the normalised offsets to 2RXS agree between the footprints to within about 1\% at the cut. In the western footprint, sources are in addition matched against eRASS1 within $4\arcmin$. Of the 61\,428 sources, 27\,807 have no association in these catalogues. Unless stated otherwise, fractions below are corrected for chance matches, estimated by repeating the search with positions displaced by $30\arcmin$. Adding eRASS1 in the western footprint reduces the western unmatched fraction from 44.6 to 9.0\% before chance correction. After chance correction the western unmatched fraction is 10.0\%, consistent with the 12.5\% implied by the test-region reliability (Sect.~\ref{sec:validation}). It is slightly lower because this sample also includes the training and validation regions and accepts counterparts from the other catalogues. The held-out value of 87.5\% is therefore adopted as the reference purity of the western catalogue. Here ``spurious'' means a detection without a point-like X-ray counterpart. Pure noise peaks are rare in the smoothed input image, so these are mostly fragments of diffuse or extended emission detected as point sources. Blends of several real sources, on the other hand, usually retain an eRASS1 counterpart within the matching radius and therefore count as real, although their fluxes are biased high (Sect.~\ref{sec:products}). 

Considering that sources without known X-ray counterparts in eastern half of the sky are potentially real new sources unknown to this day, I also create a  supplementary list containing 13\,983 unmatched candidates in the eastern footprint. Those are, i.e. sources with no counterpart in any of the public ROSAT, XMMSL3, 5XMM, or ART-XC lists and for which no eRASS1 reference is available. Those represent roughly 1\% of the brightes X-ray sources actually detected by eROSITA in eastern part of the sky and provide a first glimpse on ``the other'' half of the survey for general scientific community. 

Their spurious probabilities are estimated from S/N and morphology in the western footprint and adjusted using the flux-dependent probability that a real source lacks an external counterpart. Summing the real-source probabilities gives approximately 10\,800 candidates expected to be real. Here purity is the expected fraction of real sources in a selection, $1-\langle P_{\rm spur}\rangle$, so it trades against the number of selected candidates (Table~\ref{tab:purity}). Taking all candidates gives a purity of 77\%; a cut at $P_{\rm spur}\leq0.1$ selects 2\,680 sources with a model-estimated purity of 94.1\%, and $P_{\rm spur}\leq0.05$ gives 1\,006 sources at 98.5\%. Purity is limited mainly by S/N and morphology: point-like candidates are several times cleaner than extended or saturated ones at the same S/N, and P(spurious) falls from about 0.33 below S/N 3.9 to about 0.07 above S/N 7 for point-like sources. Users should therefore choose the cut to match their tolerance to contamination rather than treat the full list as a source catalogue.

\begin{table}[t]
\caption{Expected yield and purity of eastern-footprint candidates as a function of the $P_{\rm spur}$ cut.}
\label{tab:purity}
\centering
\begin{tabular}{lrrrr}
\toprule
$P_{\rm spur}\leq$ & $N$ & Real & Spurious & Purity (\%) \\
\midrule
1 (all) & 13\,983 & 10\,809 & 3\,174 & 77.3 \\
0.5 & 13\,395 & 10\,560 & 2\,835 & 78.8 \\
0.3 & 10\,127 & 8\,503 & 1\,624 & 84.0 \\
0.2 & 6\,956 & 6\,136 & 820 & 88.2 \\
0.1 & 2\,680 & 2\,523 & 157 & 94.1 \\
0.05 & 1\,006 & 991 & 15 & 98.5 \\
\bottomrule
\end{tabular}
\tablefoot{Expected real and spurious numbers are sums of $1-P_{\rm spur}$ and $P_{\rm spur}$ over the selection. They are model estimates transported from the western footprint, not measured purities; the held-out check suggests contamination is underpredicted by about 5\%.}
\end{table}

These estimates remain provisional: an auxiliary random-half calibration check predicts 1\,462 spurious sources where 1\,535 are inferred, underpredicting contamination by about 5\%. Applying western-derived calibration to the eastern footprint also assumes comparable external-counterpart selection at fixed flux, which the flux-scale checks alone cannot establish. Moreover, the spurious fraction transported to the eastern footprint does not vary smoothly with flux: above $2\times10^{-13}$\,\fluxunit\ the bin values range from below 0.01 to 0.24. Each bin estimate is a difference between the observed unmatched fraction and the expected fraction of real sources without external counterparts, $q(F)$, divided by $1-p_c-q(F)$, where $p_c\approx3\%$ is the chance-match rate; this denominator is only $\approx0.3$ for the faintest bins. Small errors in $q(F)$, or real differences in external-catalogue depth between the two halves, such as the deep ROSAT exposure around the north ecliptic pole in the eastern half, are therefore amplified. The total number of spurious candidates is more robust than its distribution in flux, and the yields in individual flux ranges below should be considered indicative. The total does, however, depend on the eastern flux scale, through $q(F)$: shifting eastern fluxes by the $\pm3$\% uncertainty of the flux-scale correction changes the expected number of real candidates from 10\,800 to between 10\,200 and 11\,500. In the western footprint the spurious fraction rises again to 0.15--0.26 above $6\times10^{-13}$\,\fluxunit, reflecting extended structures detected as point sources.

To summarise, of the 13\,983 candidates in the eastern footprint about 10\,800 (10\,200--11\,500 allowing for the flux-scale uncertainty) are expected to be real X-ray sources absent from ROSAT and other public X-ray catalogues, and about 3\,200 (23\%) are expected to be spurious, mostly fragments of extended emission (Table~\ref{tab:summary}). Since external counterparts of the remaining 16\,451 eastern sources confirm that they are real, about 3\,200 of all 30\,434 eastern sources outside the masks, i.e. 10\%, are expected to be spurious. The overall eastern purity of $\approx90\%$ is thus comparable to the 87.5\% measured in the western test regions, as expected if the image quality is similar in both halves. Allowing for the $\sim5\%$ underprediction of contamination above, the expected number of real new sources is $\approx10\,600$--$10\,800$. About 4\,500 of the expected real sources (from 5\,847 candidates, 78\% purity) lie above the $3.12\times10^{-13}$\,\fluxunit\ 50\% completeness limit, and about 460 (from 557 candidates) above $10^{-12}$\,\fluxunit. Some 3\,853 candidates are extended or saturated, with an expected 2\,500 real sources among them; at the bright end these are dominated by diffuse structure and saturated blobs rather than undiscovered point sources (all 73 unmatched sources above $10^{-11}$\,\fluxunit\ in the full catalogue are saturated). Point-like candidates with $P_{\rm spur}\leq0.1$, 2\,563 sources with about 2\,400 expected to be real, are therefore the most suitable starting point for follow-up. ``Real'' here means a genuine X-ray source as calibrated against eRASS1; no optical or multiwavelength identification has been attempted.

\section{Data products and limitations}
\label{sec:products}
The data release at \zenodo\ contains the catalogue in FITS and CSV format, with sky positions, positional errors, four calibrated fluxes and their bounds, detection significance, saturation and morphology indicators, edge and remnant flags, and eRASS1 associations. The training/validation/test assignment is retained so that in-sample sources can be distinguished from test sources. The release also includes the 13\,983 eastern candidates without external counterparts, with their spurious-source probabilities, in CSV and Parquet format, the three all-sky HEALPix maps described below, the input image, the run record, and full column documentation. The pipeline and its configuration are available at \github.

Per-tile products are not part of the release, as the pipeline regenerates them deterministically from the input image. They provide the data, fitted source-plus-background model, background, source-subtracted image, residual, and validity information. Originally clipped pixels in the source-subtracted tiles are assigned a fitted background estimate and recorded in a per-channel \texttt{MODELLED} plane. Such pixels are counterfactual model values, not recovered measurements. The all-sky HEALPix products use Galactic coordinates, RING ordering, and $N_{\rm side}=1024$, corresponding to approximately $3.4\arcmin$ sampling, which is finer than the effective image PSF.
All three HEALPix products contain approximate count-rate surface brightness in $\mathrm{ct\,s^{-1}\,arcmin^{-2}}$. Their R, G, and B columns correspond to equivalent rates in the eRASS1 DR1 HiPS\footnote{\url{https://erosita.mpe.mpg.de/dr1/AllSkySurveyData_dr1/HalfSkyMaps_dr1/}; these maps cover the western half at much higher resolution.} bands 0.2--0.5, 0.5--1.0, and 1.0--2.0\,keV, respectively. I smooth the reference HiPS maps to the image resolution and fit an affine relation, $S_c=a_c I_c+b_c$, to faint, unclipped pixels in each reference-footprint tile, where $I_c$ is the linearised image intensity. The median tile coefficients define the conversion applied over the full sky, and the scatter of fitted slopes describes its spatial variation. The coefficients are fitted in the western half, the only one covered by the HiPS maps, and are applied to the eastern half unchanged. Across the central meridian, the eastern half of the image is 4, 5, and 8\% brighter than the western half in the R, G, and B channels, so eastern surface brightnesses may be overestimated by similar amounts. The median tile correlations are 0.74, 0.91, and 0.91 in the three channels. Band mismatch and spatial variation limit this calibration, especially in the softest channel. The FITS headers record the units and conversion coefficients. These map values are count-rate densities; the catalogue fluxes are separately calibrated in energy-flux units.

Appendix~\ref{app:maps} compares the calibrated and source-subtracted renderings at a common stretch. The latter retains unresolved sources and incompletely modelled extended emission. It also contains model-filled clipped cores; the HEALPix format does not carry the tile-level \texttt{MODELLED} mask. An additional display version interpolates the missing rendered footprint and smooths the complete map with a $10.7\arcmin$ FWHM Gaussian. These products are useful for inspection of large-scale structure but do not constitute a validated, source-free measurement of diffuse emission.

The principal catalogue limitations follow directly from the input: quantisation, saturation, confusion, and a spatially varying empirical selection function. Extended sources have no general morphology model, and blended sources can inherit a flux scale tied to only their brightest counterpart. The learned flux mapping may perform poorly for unusual spectra. The test results are conditional on global registration, fixed empirical halo safeguards, and the matched reference sample. Candidate-probability estimates have the additional qualifications given above. The run record identifies the code revision, locked environment, and checksums of the input catalogues and of the HiPS cut-outs used for the map calibration. These inputs are public and are not redistributed; the checksums allow users to verify that a re-run uses the same inputs.

\section{Conclusions}
The press-kit image released to celebrate the completion of the first eROSITA all-sky survey was intended to capture imagination of general public and scientists alike, and succeeded in this. It is amazing, however, that same artistically edited image retains enough information to support creation of an X-ray catalogue comparable in sensitivity with the best we had before eROSITA, i.e. ROSAT all sky survey. The ePNGCAT catalogue presented here contains 63\,260 sources over 99.54\% of the sky is empirically validated against eRASS1 catalogue for the western half of the Galactic sky and provides the first glimpse onto the eastern half, so far hidden from general scientific community. 

In the downstream spatial test regions, ePNGCAT yields association reliability is 87.5\% for known sources, with estimated median positional offset of $72\arcsec$ and 50\% completeness estimated at $3.12\times10^{-13}$\,\fluxunit. A restricted flux-validation sample yields a broad-band multiplicative scatter of 27\% against eRASS1, which can be considered very low considering that public image was deliberately degraded before publication. 
External comparisons support the approximate flux scale while revealing the effects of differing epochs, bands, and source selection effects.
In the eastern Galactic half, which lacks any public eROSITA data, 13\,983 sources have no counterpart in ROSAT or other public X-ray catalogues; about 10\,800 of these are expected to be real new X-ray sources so far unknown to general community, with about 3\,200 (23\%) spurious contamination. Including the externally matched sources, the estimated overall purity of the eastern catalogue, $\approx90\%$, is comparable to the western test-region reliability. A high-purity subset of about 2\,700 candidates ($P_{\rm spur}\leq0.1$, $\approx94\%$ expected purity) is provided as potential candidates for X-ray follow-up while current generation of X-ray telescopes remains operational. 

These results make the catalogue suitable for exploratory associations, approximate flux estimates, and candidate selection with explicit quality cuts. Precision source classification or population inference requires higher-resolution data and a better characterised selection function. Releasing the pipeline together with the catalogue will make the empirical choices inspectable and allow the analysis to be repeated as improved inputs become available. Alongside the catalogue also calibrated healpix maps for the three bands, and same maps with contribution of the point sources removed are released to facilitate studies of extended X-ray emission on large scale such as eROSITA bubbles and studies of X-ray background. 

\begin{acknowledgements}
This work is based on data from eROSITA, the soft X-ray instrument aboard SRG, a joint Russian-German science mission supported by the Russian Space Agency (Roskosmos), in the interests of the Russian Academy of Sciences represented by its Space Research Institute (IKI), and the Deutsches Zentrum für Luft- und Raumfahrt (DLR). The SRG spacecraft was built by Lavochkin Association (NPOL) and its subcontractors, and is operated by NPOL with support from the Max Planck Institute for Extraterrestrial Physics (MPE). The development and construction of the eROSITA X-ray instrument was led by MPE, with contributions from the Dr. Karl Remeis Observatory Bamberg \& ECAP (FAU Erlangen-Nuernberg), the University of Hamburg Observatory, the Leibniz Institute for Astrophysics Potsdam (AIP), and the Institute for Astronomy and Astrophysics of the University of Tübingen, with the support of DLR and the Max Planck Society. The Argelander Institute for Astronomy of the University of Bonn and the Ludwig Maximilians Universität Munich also participated in the science preparation for eROSITA.
This research has made use of data obtained from the 5XMM serendipitous source catalogue compiled by the XMM-Newton Survey Science Centre, the XMM2ATHENA project and in collaboration with the XMM-Newton SOC.
\end{acknowledgements}

\bibliographystyle{aa}
\bibliography{references}

\appendix
\section{Bright-source residuals}
\label{app:bright}
\begin{figure*}[p]
\centering
\includegraphics[width=0.80\textwidth,height=0.76\textheight,keepaspectratio]{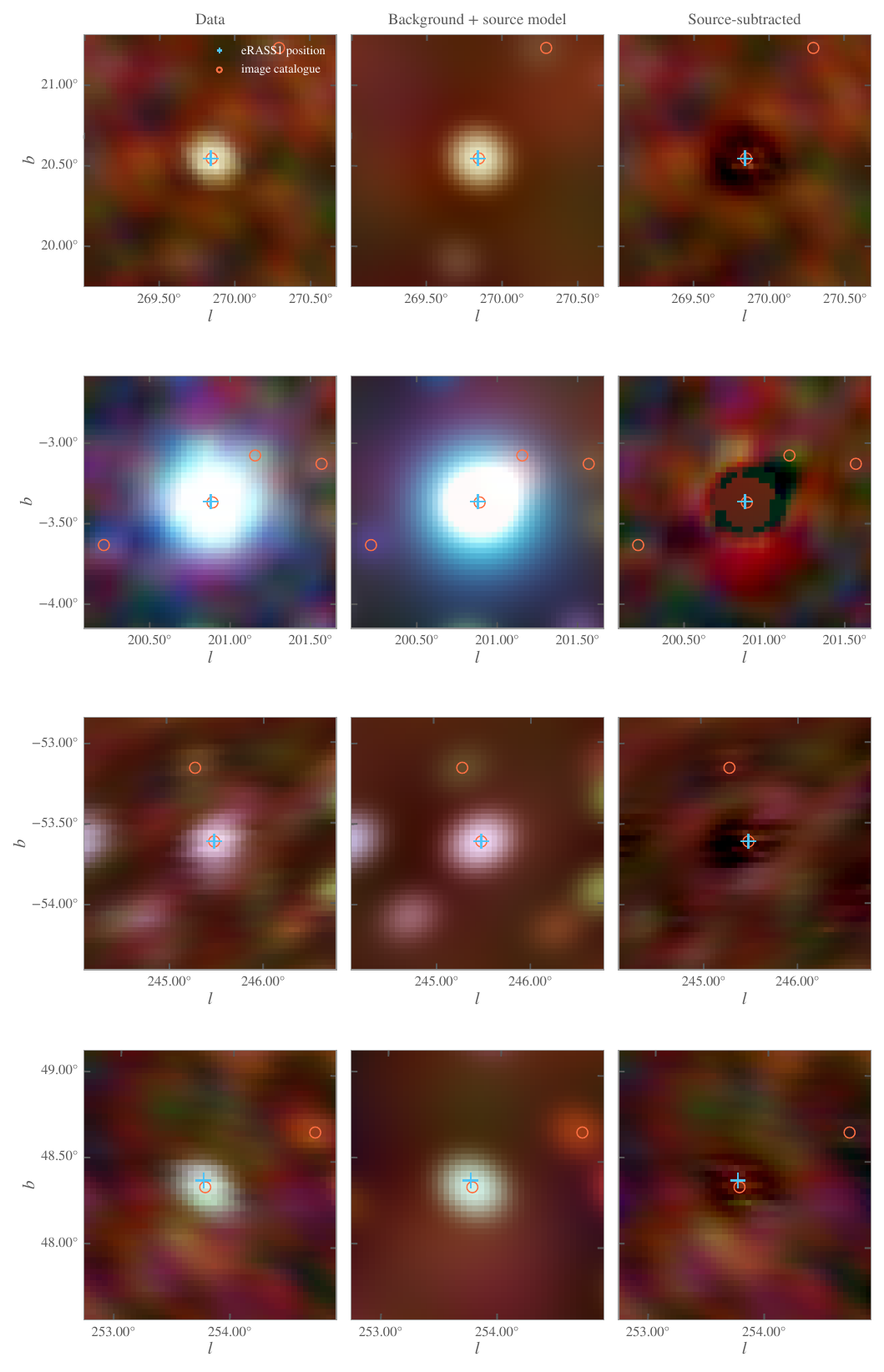}
\caption{Examples of source modelling in the tile products. Columns show the input, fitted background plus source model, and source-subtracted image in the same display units. Rows show, from top to bottom, an unsaturated point-like reference source, a saturated point-like source, an extended reference source, and a blend. The cyan cross marks the selected eRASS1 reference position; orange circles mark image-catalogue sources. Axes give Galactic longitude and latitude. The panels illustrate residuals and association ambiguities; the labels describe the reference-source selection and do not assert that every fitted component has the corresponding physical morphology.}
\label{fig:bright}
\end{figure*}

Figure~\ref{fig:bright} illustrates why source recovery cannot be characterised by a flux threshold alone. An isolated unsaturated source can be represented approximately by the local profile model. A clipped core provides much less amplitude information; an extended object leaves emission outside the fitted core; and a blend can produce one catalogue entry for several reference sources. The displayed examples are qualitative checks of the tile products. Residual structure, including deficits around some fitted cores, remains visible and should not be treated as a measurement of diffuse sky brightness.

\section{Calibrated all-sky map products}
\label{app:maps}
\begin{figure*}[p]
\centering
\includegraphics[width=0.82\textwidth]{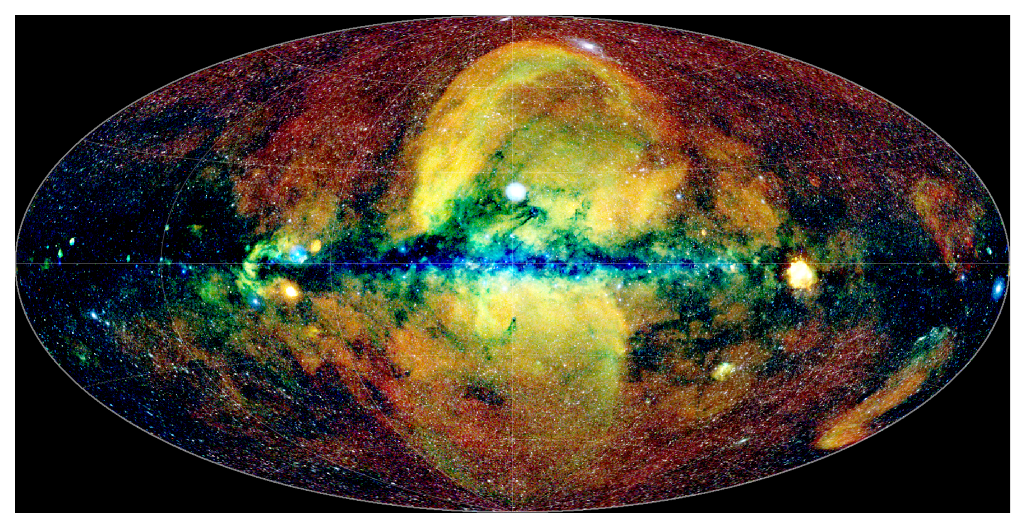}\\[2mm]
\includegraphics[width=0.82\textwidth]{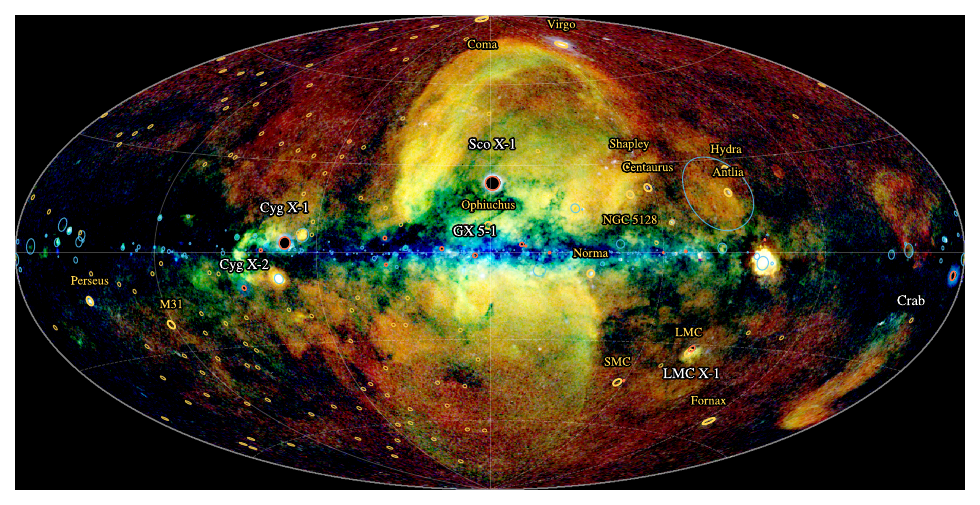}
\caption{Calibrated HEALPix image (top) and source-subtracted, gap-filled companion rendering (bottom), in Galactic coordinates. Both use the same per-channel display limits and asinh stretch; colours correspond to the three image channels after empirical calibration against the eRASS1 HiPS bands. The subtraction removes fitted source components, but unresolved sources and extended structures remain. Cyan outlines identify SNRcat remnants; yellow outlines and labels identify broad or named extended objects. Orange outlines identify additional bright-source masks applied for this plot. Clipped-core pixels replaced by model backgrounds and interpolated footprint gaps are not recovered observations. The displayed map does not propagate the tile-level modelling mask.}
\label{fig:maps}
\end{figure*}

The common display stretch in Fig.~\ref{fig:maps} allows the removal of compact structure to be inspected without independently rescaling the source-subtracted image. Remaining large structures include diffuse Galactic emission, remnants, and extended extragalactic sources. Selected bright-source regions are additionally masked for this display. These masks, the interpolation of the small footprint gap, and the model replacement of clipped pixels must be distinguished from actual measurements of source-free emission.
\end{document}